# MC CDMA PAPR REDUCTION TECHNIQUES USING DISCRETE TRANSFORMS AND COMPANDING

B.Sarala[1] and D.S.Venkateswarulu[2]

[1]Department of ECE, M V S R Engineering College, Hyderabad
b.sarala@rediffmail.com
[2]Department of ECE, Progressive Engineering College, Cheekati Mamidi, HMDA, Hyderabad.
dsv4940@gmail.com

## ABSTRACT

*High Peak to Average Power Ratio (PAPR) of the transmitted signal is a serious problem in multicarrier modulation systems. In this paper a new technique for reduction in PAPR of the Multicarrier Code Division Multiple Access (MC CDMA) signals based on combining the Discrete Transform either Discrete Cosine Transform (DCT) or multi-resolution Discrete Wavelet Transform (DWT) with companding is proposed. It is analyzed and implemented using MATLAB. Simulation results of reduction in PAPR and power Spectral Density (PSD) of the MC CDMA with companding and without companding are compared with the MC CDMA with DCT and companding, DWT and companding systems. The new technique proposed is to make use of multi-resolution DWT in combination with companding in order to achieve a very substantial reduction in PAPR of the MC CDMA signal.*

## 1. INTRODUCTION

In broadband wireless communications high bit rate transmission is required for high quality communications. The most important objective of fourth generation systems is to take care of the severe Inter Symbol Interference (ISI) resulting from high data rates. MC CDMA system is the combination of Orthogonal Frequency Division Multiplexing (OFDM) and Code Division Multiple Access (CDMA) and reaps the benefits of both the techniques. In MC CDMA, data symbols consisting of modulated bits are spread by spreading codes such as Pseudo Noise codes (PN), Gold codes, and Walsh codes and then mapped into subcarriers of an MC CDMA modem data symbol which is spread across frequency domain [1]. MC CDMA is a very attractive technique for high speed data transmission over multipath fading channels. The PAPR problem is one of the most important issues for developing multicarrier transmission systems. MC CDMA is widely used in broadband networks such as Long Term Evaluation (LTE) and broadband communication networks [2]. However, MC CDMA systems have the inherent problem of a high PAPR, which causes poor power efficiency or serious performance degradation in the transmitted signal. This brings disadvantages like complexity of ADC and DAC, reduced power efficiency, and high Bit Error Rate (BER), consumption of more power. High power amplifiers are required which results in increased cost component. Thus, if we reduce PAPR, we shall obtain reduced complexity of Analog to Digital Converter (ADC) and Digital to Analog Converters (DAC), improved signal to noise ratio and Bit Error Rate (BER) [3]. To reduce the PAPR, many techniques are proposed [4].

The first one is the signal distortion technique, which introduces distortion to signals and causes degradation in the performance including clipping, windowing, peak cancelling or companding. In companding technique, compression in transmitter and expanding in receiver has been proposed by Wang et al [5, 6]. Clipping is simple and effective and causes In-Band-distortion and increased BER. The companding transforms' performance is better and reduces distortion than to that of the clipping. Another proposal by Yuan Jiang [5] is an algorithm that uses the special airy function and is able to provide an improved Bit Error Rate (BER) and minimized Out of Band Interference (OBI) in order to reduce PAPR effectively.

Different coding techniques are proposed for signal scrambling techniques which can be further classified into: schemes with explicit side information including linear block codes, selective mapping (SLM), and partial transmit sequence (PTS), interleaving schemes. There is an ease of modification, increased overhead and increased search complexity with increased data loss. The various schemes proposed without side information include block coding, Hadamard transform method, dummy sequence insertion method etc. Signal scrambling technique may not be affecting the system performance but it has an overhead of increased complexity and needs to perform exhaustive search to find best codes and to store large lookup tables for encoding and decoding. It does not support error correction [7, 8].Error control selective mapping is effective and there is no need for side information but, complexity is increased. Other schemes are proposed using separation of complex baseband signal for all modulations, which improve performance and result in less complexity [9].

Zhongpeng Wang suggested combining DCT and companding for PAPR reduction in OFDM signals [10]. A DCT may reduce the PAPR of multicarrier modulation signal, but does not increase the BER of system. In this paper, a new technique is proposed that makes use of DCT/DWT in combination with companding in order to achieve a very substantial reduction in PAPR, and PSD of the MC CDMA signal. In this scheme, in the first step, the data is transformed by a Discrete Cosine Transform (DCT) or Discrete Wavelet Transform into new modified data. In the second step, this technique utilizes the companding technique to further reduce the PAPR of the MC CDMA signal. The DCT/DWT may reduce PAPR of an MC CDMA signal, but does not increase the BER of system.

This paper computed and compared PAPRs, and PSDs of MC CDMA original, MC CDMA with companding, MC CDMA with DCT and companding, MC CDMA with DWT and companding. Simulation results show that the PAPRs of DWT with companding based MC CDMA system has low PAPR, lower side lobe and higher bandwidth efficiency. This technique reduces PAPR and PSD without degradation of BER performance.

The rest of the paper is organized as follows: Section 2 describes MC CDMA system and PAPR models are discussed. In section 3 related works are discussed. In section 4 proposed system model of MC CDMA either with Discrete Transform or wavelet transform with companding is discussed. Section 5 details the proposed technique with the help of a suitable algorithm. Computer simulations are presented in section 6. Finally, conclusions are listed in section 7.

## 2. MC CDMA SYSTEM & PAPR REDUCTION

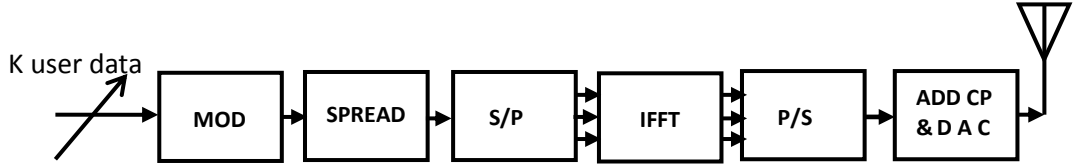

Figure 1. Block Diagram of MC CDMA Transmitter

The MC CDMA transmitter configuration for the $k^{th}$ user is shown in figure 1. After modulation, user data is fed to the spreader. The $n^{th}$ symbol of the $k^{th}$ user in the $i^{th}$ block by user specific spreading code $c_k(t) = [c_1^k, c_2^k, ----, c_{Nc}^k]$ is the spreading factor of the $k^{th}$ user is fed to serial to parallel converter and then Inverse Fast Fourier Transform (IFFT) in the frequency domain. MC CDMA uses Inverse Fast Fourier Transform (IFFT) to divide the bandwidth into orthogonal overlapping subcarriers; the $n^{th}$ data symbol for $k^{th}$ user, $a_k(n)$ is spread by user k's corresponding spreading code vector $C_k$. Each of the $N_c$ subcarriers is modulated by a single chip.

The data is converted back into serial data before cyclic prefix or guard interval is inserted to combat ISI. Finally the signal is fed to Digital to Analog converter for transmission (DAC) through Rayleigh fading channel with Additive White Gaussian Noise (AWGN). The transmitted signal is represented as

$$p_k(t) = \sum_{s=0}^{N_c-1} a_k(n) c_k(s) exp(jw_m t), \qquad (n-1)T_c \leq t \leq nT_c \qquad (1)$$

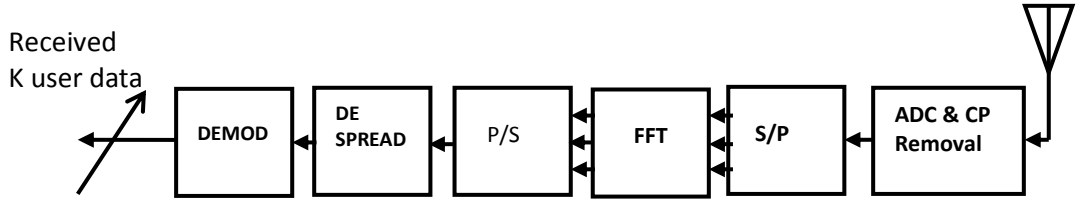

Figure 2. Block Diagram of MC CDMA Receiver

The MC CDMA receiver model for $k^{th}$ user is shown in figure (2). The received signal is first down converted, and the cyclic prefix or guard interval is removed. Then, the data is fed to serial to parallel converter. After that, the signal is transformed using FFT and fed to dispreading and demodulation blocks [11].

For $Tc=T_s$ (the chip duration which equals to the symbol duration), the combined signal for all k users is at the receiver can be denoted as

$$q_k(t) = \sum_{s=0}^{N_c-1} \sum_{k=1}^{K} a_k(n) c_k(s) exp(jw_m t), \qquad (n-1)T_c \leq t \leq nT \qquad (2)$$

## 2.1 PAPR problem of MC CDMA signals

MC CDMA signal consists of n data symbols transmitted over $N_c$ subcarriers. Let

P={ $P_k$, k=0,1,2,--,$N_c$-1} be a block of n data symbols and each symbol modulating a set of subcarriers {$f_k$, k=0,1,- -,$N_c$-1}. The $N_c$ subcarriers are orthogonal, that is where $f_k=k\Delta f$, $\Delta f=1/N_c T$ and T is the actual symbol period.

In, general the PAPR of the MC CDMA signal p (t) is defined as the ratio between maximum instaneous power and its average power during the MC CDMA signal.

$$PAPR = \frac{\max[|p(t)|^2]}{E[|p(t)|^2]} \qquad (3)$$

Where E [.] denotes expectation and complementary cumulative distribution function for MC CDMA signal can be written as CCDF = probability (PAPR> $P_0$), where $P_0$ is the Threshold [12].

PAPR of MC CDMA signal is mathematically defined as

$$PAPR = 10\, log_{10} \frac{\max[|p(t)|^2]}{\frac{1}{T}\int_0^T |p(t)|^2\, dt} \quad dB \qquad (4)$$

It is easy to manipulate the above equation by decreasing the numerator max [|p (t)|²] or increasing the denominator E[|p(t)|²] or both.

## 3. RELATED WORK

Zhongpeng Wang proposed Combined DCT and companding technique for PAPR reduction in Orthogonal Frequency Division Multiplexing (OFDM) signals. In his scheme, as a first step, the data is transformed by a Discrete Cosine Transform/Discrete Wavelet Transform (DCT/DWT) into new modified data. In the second step, the proposed technique utilizes the companding technique to further reduce the PAPR of the OFDM signal. The DCT may reduce PAPR of an OFDM signal, but does not increase the BER of system. Haixia Zhang [13, 14] proposed a study on the PAPRs in multicarrier modulation systems with different orthogonal bases, one Fourier and six wavelet bases. Zhang then introduced a novel threshold based PAPR reduction method in Wavelet based Multicarrier Modulation (WMCM) systems. This method works very effectively in WMCM systems. The distortion of the wavelet transform caused by the threshold is measured in terms of Mean Square Error (MSE).

Earlier we proposed the technique for improving Bit Error Rate (BER) and Signal to Noise Ratio (SNR) of MC CDMA considering three different spreading sequences in presence of Additive White Gaussian Noise (AWGN) and Rayleigh fading channels. Bit Error Rate (BER) of MC CDMA transmission system depends strongly on Multiple Access Interference (MAI) due to cross correlation properties of applied spreading codes. The spreading codes for MC CDMA like Walsh codes, Gold codes, and Maximal length Pseudo Noise (PN) codes are used in order to minimize the BER, and to reduce MAI [15, 16].

This paper deals with the use of DCT/DWT in combination with companding in order to achieve a very substantial reduction in PAPR of the MC CDMA signal. In the proposed scheme, in the first step, the data is transformed by a Discrete Cosine Transform (DCT) or Discrete Wavelet Transform into new modified data. In the second step, the proposed technique utilizes the companding technique to further reduce the PAPR of the MC CDMA signal. The DCT may reduce PAPR of an MC CDMA signal, but does not increase the BER of system. This paper has implemented the same proposed techniques to reduce the PAPR and PSD for MC CDMA system.

In the present paper, we first compare the PAPRs of MC CDMA original, MC CDMA with companding, MC CDMA with DCT and companding, MC CDMA with Haar DWT and companding. Simulation results show that the PAPRs of Haar DWT with companding based MC CDMA system have low PAPR when compared with other MC CDMA systems. The power spectral density of the resultant signal has lower side lobes which minimize interference between signals. The proposed technique reduces PAPR, without degradation in BER performance, and the resultant system has less mean amplitude over conventional MC CDMA techniques.

## 4. PROPOSED MC CDMA SYSTEM

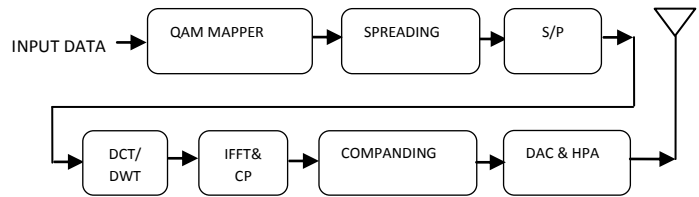

Figure 3. Block diagram of MC CDMA transmitter with DCT/ DWT & Companding

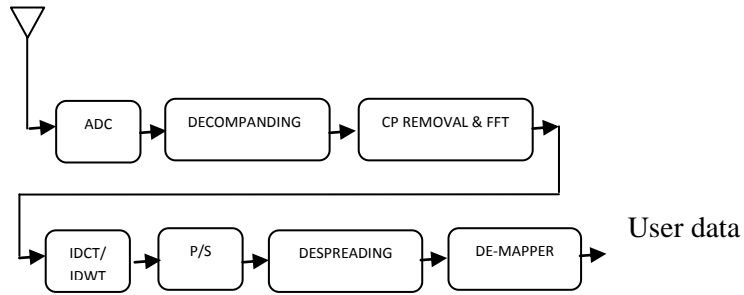

Figure 4. Block diagram of MC CDMA receiver with IDCT/IDWT & Decompanding

### 4.1 Discrete Cosine Transform

In MC CDMA system a precoding matrix m of dimension N x N is constructed which is based on Discrete Cosine Transform (DCT). P is applied to constellation symbols before IFFT to reduce the PAPR. DCT matrix m is created by using equation (5)

$$b(k) = \left\{\frac{1}{\sqrt{N}}, for\ k = 0\ ,\sqrt{\frac{2}{N}}, for\ k \neq 0\right. \tag{5}$$

The formal definition of a one dimensional DCT of length is represented as

$$P_c(K) = b(k) \sum_{n=0}^{N-1} p(n) \cos\left[\frac{\pi(2n+1)K}{2N}\right] \quad for\ K = 0, --, N-1 \qquad (6)$$

Similarly, the inverse transformation is defined as

$$p(n) = \sum_{n=0}^{N-1} P_c(K) b(k) \cos\left[\frac{\pi(2n+1)K}{2N}\right] \quad for\ n = 0, \ldots, N-1 \qquad (7)$$

The equation (6) expressed in matrix form as

$$P_c = C_N\ p \qquad (8)$$

$P_c$, p are vectors of dimension N x1 and $C_N$ is a matrix of dimension N x N. The rows or columns of the DCT matrix, $C_N$ are orthogonal vectors. We can use this property of the DCT matrix to reduce the PAPR of MC CDMA signals.

## 4.2 Discrete Wavelet Transform

The wavelet transform is a kind of technique derived from Fourier Transform wherein the localization property of the wavelet functions along with localization of frequency provides very efficient characteristics [13, 14].The main characteristic of the wavelet transform is to provide side lobes of much lower magnitude than of Fourier Transforms. This is a very important reason to have used wavelet bases to modulate symbols in Multi Carrier Modulation systems. The wavelet packet functions $w_k$ (t) (n ϵ $Z_+$) is followed by the recursive functions:

$$w_{2k}(t) = \sqrt{2} \sum_{n\ \epsilon\ Z_+} h(n)\ w_k(2t - n) \qquad (9)$$

$$w_{2k+1}(t) = \sqrt{2} \sum_{n\ \epsilon\ Z_+} g(n)\ w_k(2t - n) \qquad (10)$$

Where h (n) and g (n) = (-1)$^k$ h (L-L-1-n) stand for a pair of quadrature mirror filters (QMFs) of length L.

The discrete wavelet transform (DWT) is a type of batch processing, which analyzes a finite length time domain signal by breaking up the initial domain in two parts: the detail and approximation information. The DWT property is only few coefficients of DWT dominates the representation.

## 4.3 Companding Transform

The companding transformation is applied at the transmitter after IFFT block in order to attenuate the high peaks and amplify low amplitude of the MC CDMA signal, thus decreasing the PAPR. At the receiver, the decompanding process is applied by using the inverse companding function prior to FFT block in order to recover the original MC CDMA signal.

For the discrete MC CDMA signal is given by equation (1), the companded signal for μ law (u law) can be given by,

$$v(n) = C\, p(n) = \frac{sp(n)}{ln(1+u)p(n)} ln\left(1 + \frac{u}{s}|p(n)|\right) \tag{11}$$

Where s is the average amplitude of the signal and μ is the companding factor. The companding transform satisfy the following conditions.

$$E(|v(n)|^2) = E(|p(n)|^2) \tag{12}$$

$$|v(n)| \geq P(n), for\ |p(n) \leq s| \tag{13}$$

$$|v(n)| \leq p(n), for\ |p(n) \geq s| \tag{14}$$

At the receiver side, the received signal must be expanded by the inverse companding transform prior to Fast Fourier Transform (FFT) processing block. The expanded signal is given by

$$Q(n) = C^{-1}r(n) = \frac{sr(n)}{u|r(n)|}\left\{exp\left[\frac{|r(n)ln(1+u)|}{s}\right] - 1\right\} \tag{15}$$

## 5. A NEW PROPOSED MC CDMA SCHEME

To reduce the PAPR an MC CDMA signal, we propose a new approach i.e. the combination of a companding transform and Discrete Transform either Discrete Cosine Transform (DCT) or multi-resolution Haar Discrete Wavelet Transform (DWT) with companding for various spreading sequences like PN sequence, Gold codes, and Walsh Hadamard codes. In the proposed scheme, in the first step the data is transformed by a Discrete Cosine Transform (DCT) or Discrete Wavelet Transform into new modified data. In the second step, the proposed technique utilizes the companding technique to further reduce the PAPR of the MC CDMA signal. The input data is modulated and multiplied with spreading code and is processed with a DCT or DWT then with an IFFT block as shown in Figure (3).

The following steps for the MC CDMA as follows:

Step1: The input binary sequence b (n) is modulated and then multiplied with spreading code. The output of spreader is p is represented as

$$p = a_k(n)c_k(s).$$

Step2: The sequence p is transformed using the DCT or DWT matrix, i.e. $Q = HP$.

Step3: An $IFFT(Q)$ is applied

$$q = [q(1), q(2), \ldots, q(N)]^T.$$

Step4: A companding transform is then applied to Q .i.e. $V(n) = C\{q(n)\}$.

Step5: An inverse companding transform is applied to the received signal, r(n), i.e.

$$\hat{q}(n) = C^{-1}\{r(n)\}$$

Step6: A FFT transform is applied to the signal,

$\hat{q}(n), i.e.\ \hat{Q} = FFT(\hat{q})$, where

$\hat{q} = [\hat{q}(1), \hat{q}(2), \ldots, \hat{q}(N)]^T$

Step7: An inverse DCT or DWT transforms applied to the signal, $\hat{Q}\ i.e.\ \hat{P} = H^T \hat{Q}$

Step8: Then the signal $\hat{P}$, is dispread and de-mapped from the bit stream to get original binary sequence $\hat{b}(n)$.

## 6. SIMULATION RESULTS

Original MC CDMA, MC CDMA with companding, and MC CDMA with DCT and companding, Haar DWT and companding systems are implemented using MATLAB with the following specifications: number of symbols are 512, IFFT size is 128, and number of subcarriers are 64, spreading codes are PN codes, Gold codes, Walsh Hadamard codes and modulation used Binary Phase Shift Keying (BPSK), Quadrature Phase Sift Keying (QPSK) with µ companding values are 2, 3, 5. We can evaluate the performance of PAPR and BER using complementary cumulative distribution of PAPR of MC CDMA with DCT and companding, DWT and companding. The results are compared with original MC CDMA (without DCT or DWT or companding), MC CDMA with companding, and MC CDMA with DCT and companding, and MC CDMA with DWT and companding. In MATLAB simulation we assumed that µ is equal to u.

### 6.1 CCDF Performance

We can evaluate the performance of PAPR using cumulative distribution of PAPR of MC CDMA signal. The Complementary Cumulative Distribution Function (CCDF) is one of the most regularly used parameters, which is used to measure the efficiency of PAPR technique.

The figure 5 shows the CCDF performance of a MC CDMA original system, MC CDMA with companding, and MC CDMA with DCT and companding algorithm for PAPR reduction. The figure shows that when CCDF = $10^{-2}$ in MC CDMA with the companding method, the PAPR is reduced by 3.75dB, 5.25 dB, and 6.75 dB for values of µ is 2, 3, and 5 respectively when compared to original MC CDMA system. At CCDF = $10^{-2}$, MC CDMA with DCT and companding method, the PAPR is reduced by 5.75 dB, 6.75 dB, and 7.75 dB for values of µ is 2, 3, and 5 respectively when compared with the original MC CDMA system. The simulation results show that the MC CDMA with DCT and companding scheme resulted in about 1.33 dB PAPR reduction compared with MC CDMA with companding.

The figure 6 shows the CCDF performance of a MC CDMA original, MC CDMA with companding, and MC CDMA with DCT and companding algorithm for PAPR reduction. The figure shows that when CCDF = $10^{-2}$ in MC CDMA with the companding method the PAPR is reduced by 3.5dB, 5.15 dB, and 6.25 dB for values of µ is 2, 3, and 5 respectively when compared to original MC CDMA system. At CCDF = $10^{-2}$, MC CDMA with DCT and companding method, the PAPR is reduced by 4.5 dB, 6.0 dB, and 7.15 dB for values of µ is 2, 3, and 5 respectively when compared with the original MC CDMA system. The simulation results show that the MC CDMA with DCT and companding scheme resulted in about 0.92 dB PAPR reduction compared with MC CDMA and companding.

The figure 7 shows the CCDF performance of a MC CDMA original, MC CDMA with companding, and MC CDMA with DCT and companding algorithm for PAPR reduction. The figure shows that when CCDF = $10^{-2}$ in MC CDMA with the companding method the PAPR is reduced by 3.25dB, 5.25 dB, and 6.25 dB for values of μ is 2, 3, and 5 respectively when compared to original MC CDMA system. At CCDF = $10^{-2}$, MC CDMA with DCT and companding method, the PAPR is reduced by 5.25 dB, 6.75 dB, and 7.9 dB for values of μ is 2, 3, and 5 respectively when compared with the original MC CDMA system. The simulation results show that the MC CDMA with DCT and companding scheme resulted in about 1.12dB PAPR reduction compared with MC CDMA with companding.

The figure 8 shows the CCDF performance of a MC CDMA original, MC CDMA with companding, and MC CDMA with DCT and companding algorithm for PAPR reduction. The figure shows that when CCDF = $10^{-2}$ in MC CDMA with the companding method the PAPR is reduced by 3.5dB, 5.25 dB, and 6.5 dB for values of μ is 2, 3, and 5 respectively when compared to original MC CDMA system. At CCDF = $10^{-2}$, MC CDMA with DCT and companding method, the PAPR is reduced by 5.75 dB, 6.95 dB, and 8.25 dB for values of μ is 2, 3, and 5 respectively when compared with the original MC CDMA system. The simulation results show that the MC CDMA with DCT and companding scheme resulted in about 1.7dB PAPR reduction compared with MC CDMA with companding.

The figure 9 shows the CCDF performance of a MC CDMA original, MC CDMA with companding, and MC CDMA with DCT and companding algorithm for PAPR reduction. The figure shows that when CCDF = less than $10^{-2}$ in MC CDMA with the companding method the PAPR is reduced by 3dB, 5 dB, and 6 dB for values of μ is 2, 3, and 5 respectively when compared to original MC CDMA system. At CCDF = less than $10^{-2}$, MC CDMA with DCT and companding method, the PAPR is reduced by 4.5 dB, 5.5 dB, and 7 dB for values of μ is 2, 3, and 5 respectively when compared with the original MC CDMA system. The simulation results show that the MC CDMA with DCT and companding scheme resulted in about 1dB PAPR reduction compared with MC CDMA with companding system.

The figure 10 shows the CCDF performance of a MC CDMA original, MC CDMA with companding, and MC CDMA with DCT and companding algorithm for PAPR reduction. The figure shows that when CCDF = less than $10^{-2}$ in MC CDMA with the companding method the PAPR is reduced by 3dB, 5 dB, and 6 dB for values of μ is 2, 3, and 5 respectively when compared to original MC CDMA system. At CCDF = less than $10^{-2}$, MC CDMA with DCT and companding method, the PAPR is reduced by 5 dB, 6.5 dB, and 8 dB for values of μ is 2, 3, and 5 respectively when compared with the original MC CDMA system. The simulation results show that the MC CDMA with DCT and companding scheme resulted in about 1.83dB PAPR reduction compared with MC CDMA with companding system.

The figures 11, 12, 13 show the CCDF performance of a MC CDMA original, MC CDMA with companding, and MC CDMA with DCT and companding, MC CDMA with DWT and companding algorithm for PAPR reduction using different spreading codes. Figures show that when CCDF = $10^{-2}$ in MC CDMA with the companding method the PAPR is reduced by 3.75dB, 5.37 dB, and 6.5 dB for μ is 2, 3, and 5 when compared to original MC CDMA system.

At CCDF = $10^{-2}$, MC CDMA with DCT and companding method, the PAPR is reduced by 4.8 dB, 5.92 dB, and 7.2 dB for values of μ is 2, 3, and 5 respectively when compared with the original MC CDMA system. The simulation results show that the MC CDMA with DCT and companding scheme obtained about 0.76dB PAPR reduction compared with MC CDMA with companding system.

At CCDF = $10^{-2}$, and MC CDMA with DWT and companding method, the PAPR is reduced by 5.78dB,7dB,8.2dB for values of μ is 2, 3, and 5 respectively when compared with MC CDMA system with DCT and companding. The simulation results show that the MC CDMA with DWT and companding scheme obtained about 1.0dB PAPR reduction compared with MC CDMA with DCT and companding system.

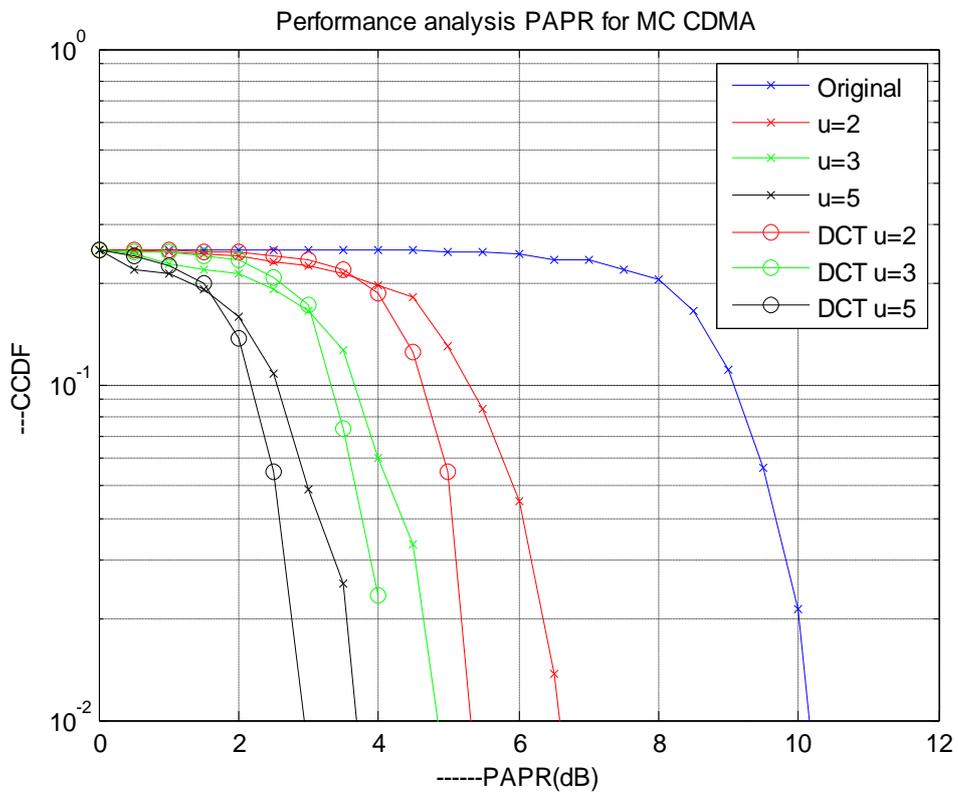

Figure 5. MC CDMA with DCT and companding using BPSK modulation random PN sequence comparisons of the CCDF of different companding factors.

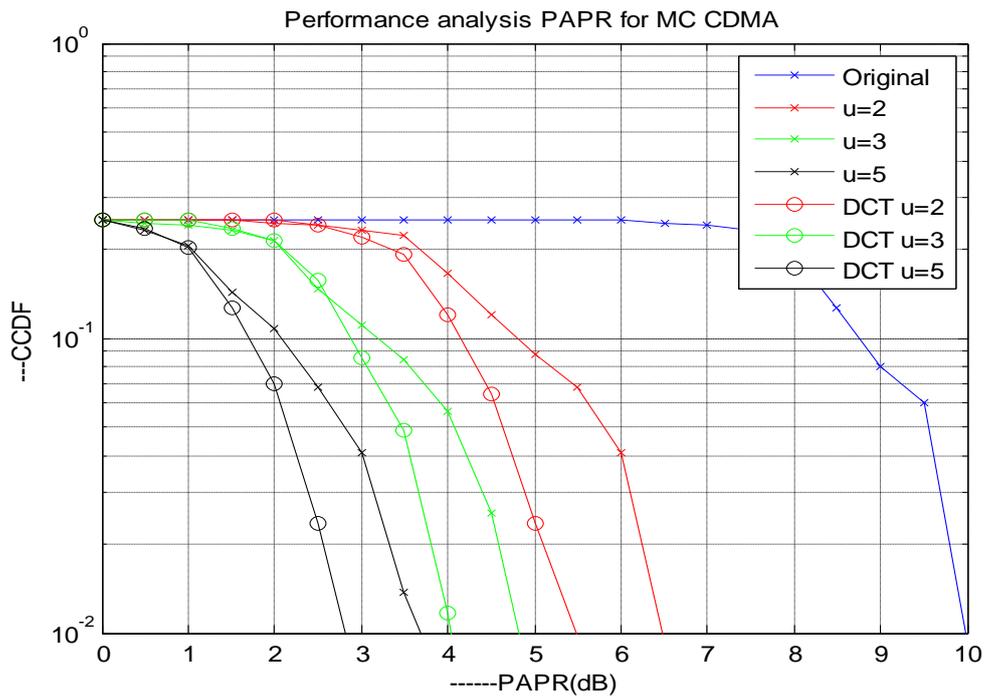

Figure 6. MC CDMA with DCT and companding using QPSK modulation and random PN sequence comparisons of the CCDF of different companding factors.

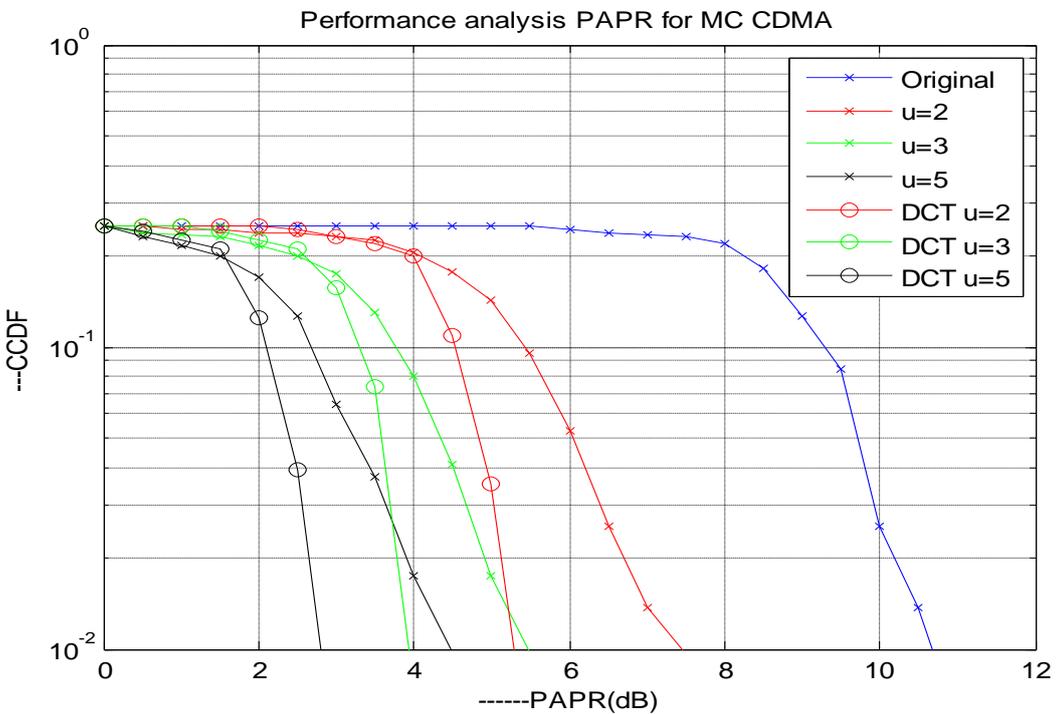

Figure 7. MC CDMA with DCT and companding using BPSK modulation and Gold code sequence comparisons of the CCDF of different companding factors.

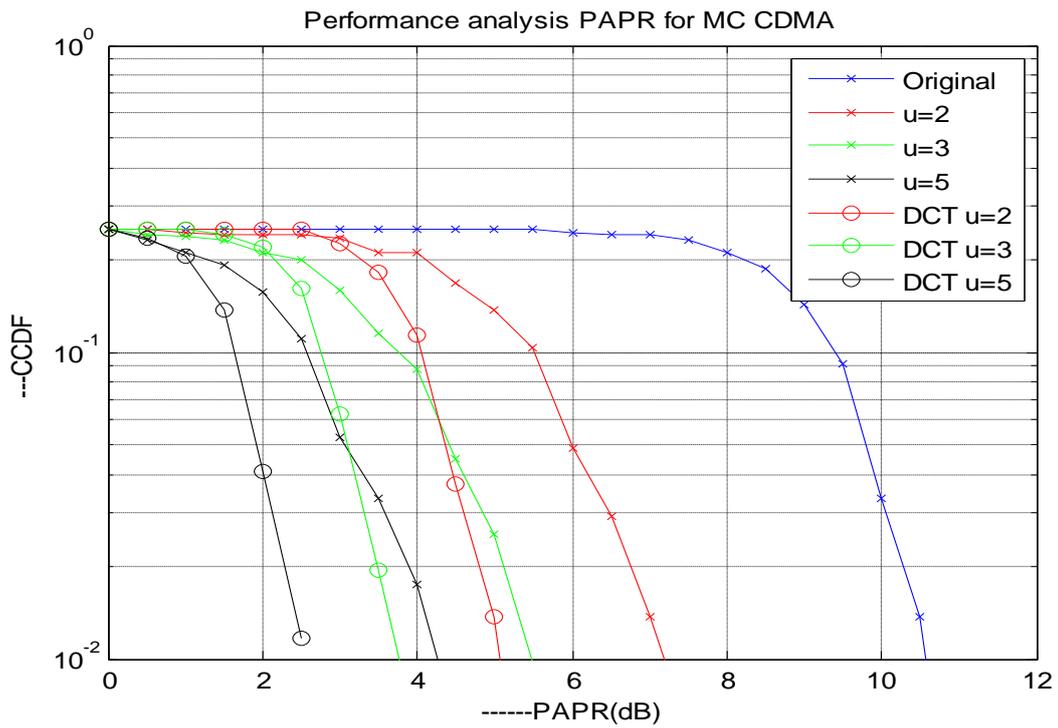

Figure 8. MC CDMA with DCT and companding using QPSK modulation and Gold code sequence comparisons of the CCDF of different companding factors.

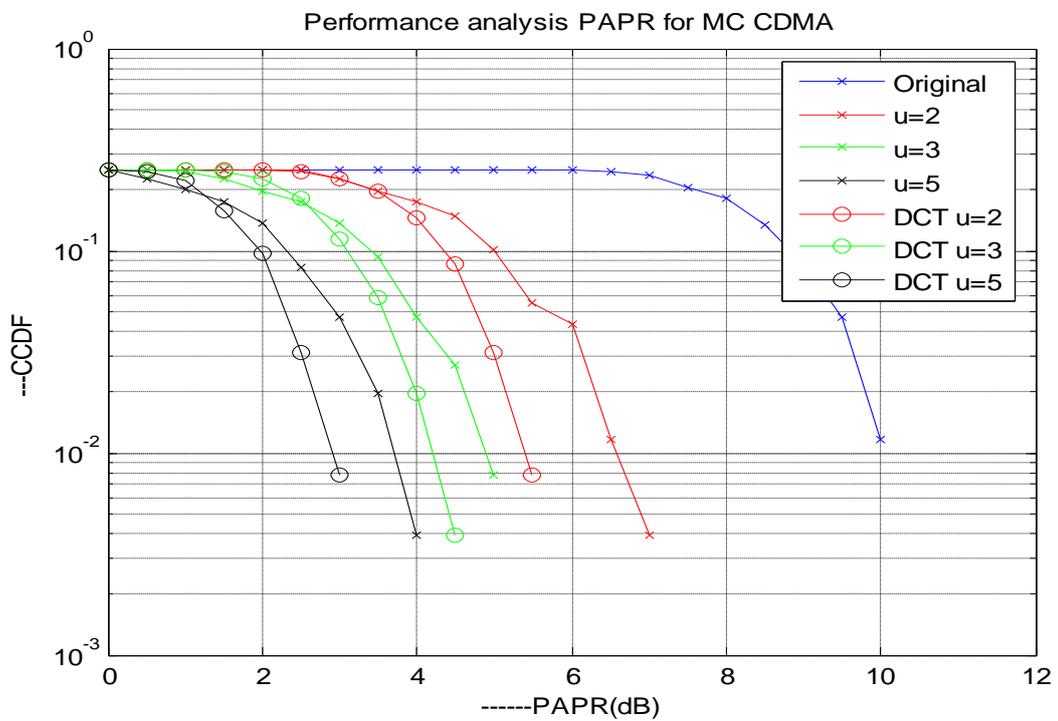

Figure 9. MC CDMA with DCT and companding using BPSK modulation and Walsh Hadamard code sequence comparisons of the CCDF of different companding factors.

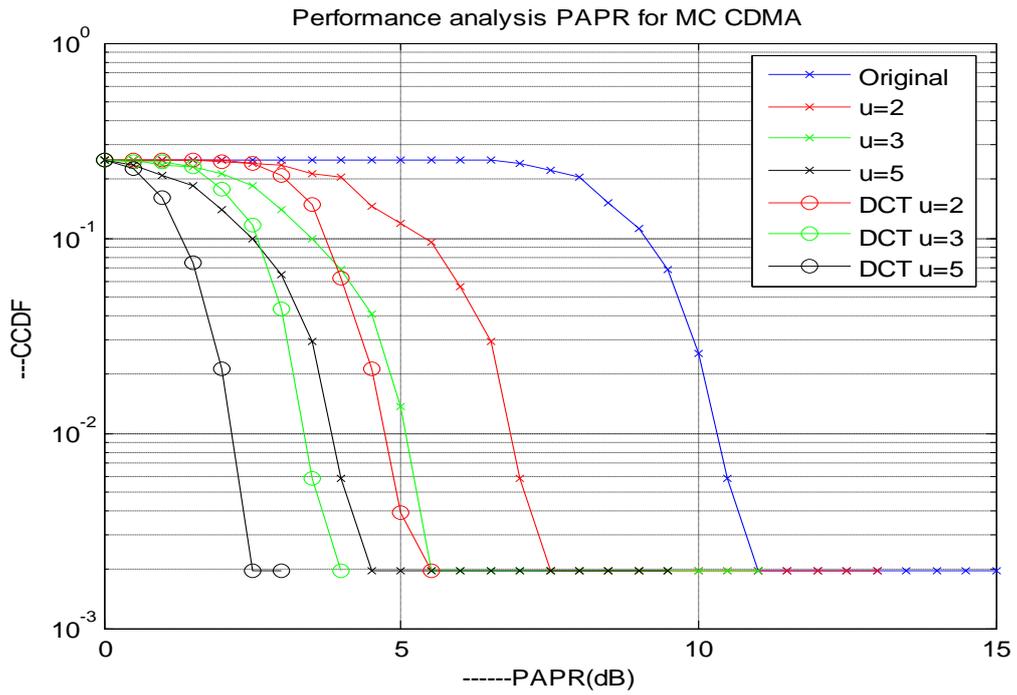

Figure 10. MC CDMA with DCT and companding using QPSK modulation and Walsh Hadamard code sequence comparisons of the CCDF of different companding factors.

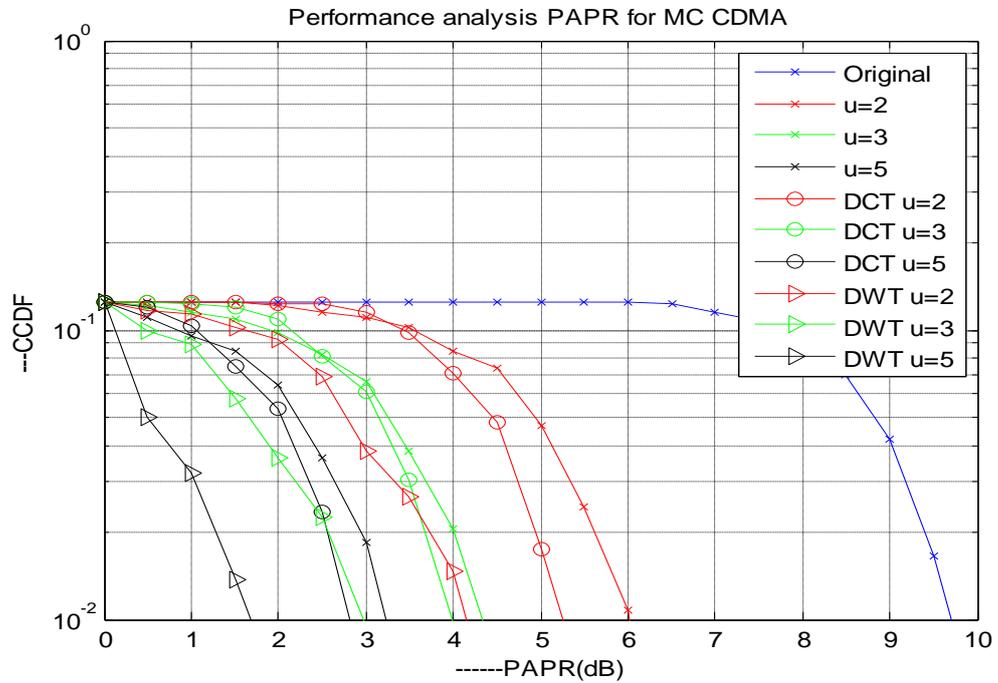

Figure 11. PAPR reduction using PN sequence and BPSK modulation with 1024 symbols.

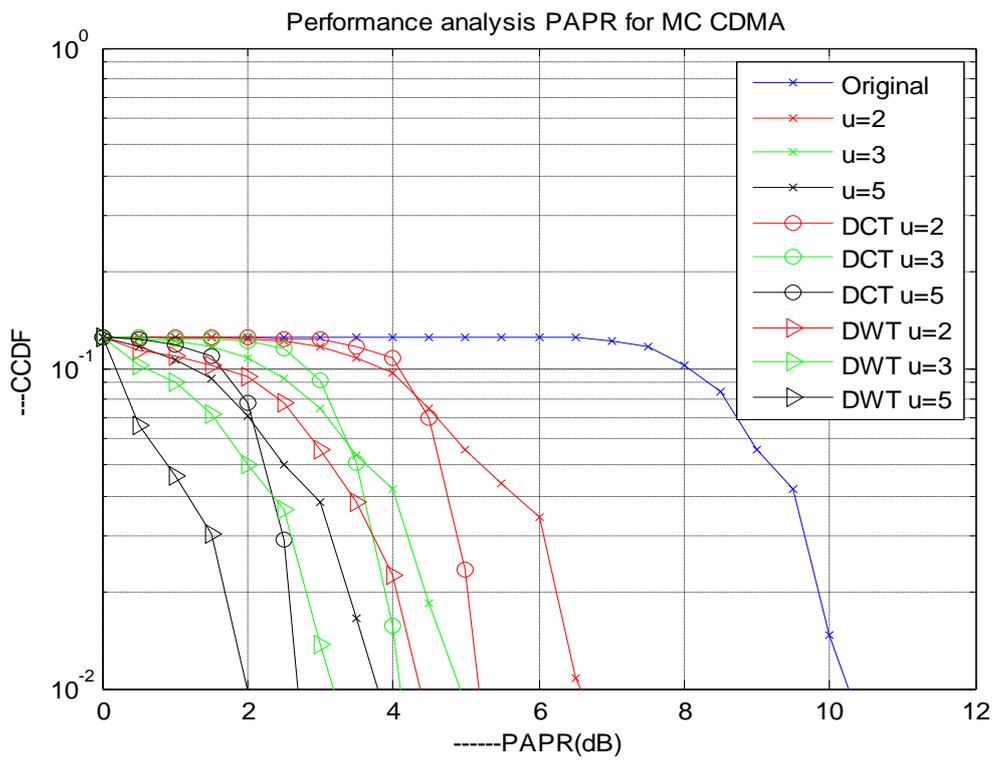

Figure 12. PAPR reduction using Gold codes and BPSK modulation with 1024symbols.

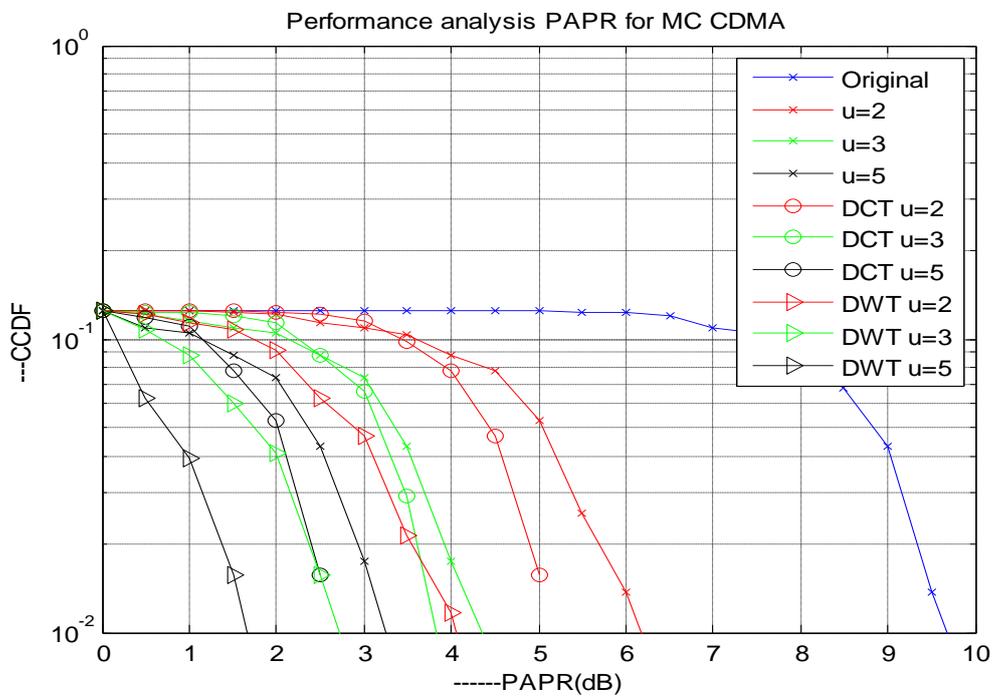

Figure 13. PAPR reduction using Walsh Hadamard codes, BPSK modulation with 1024 symbols.

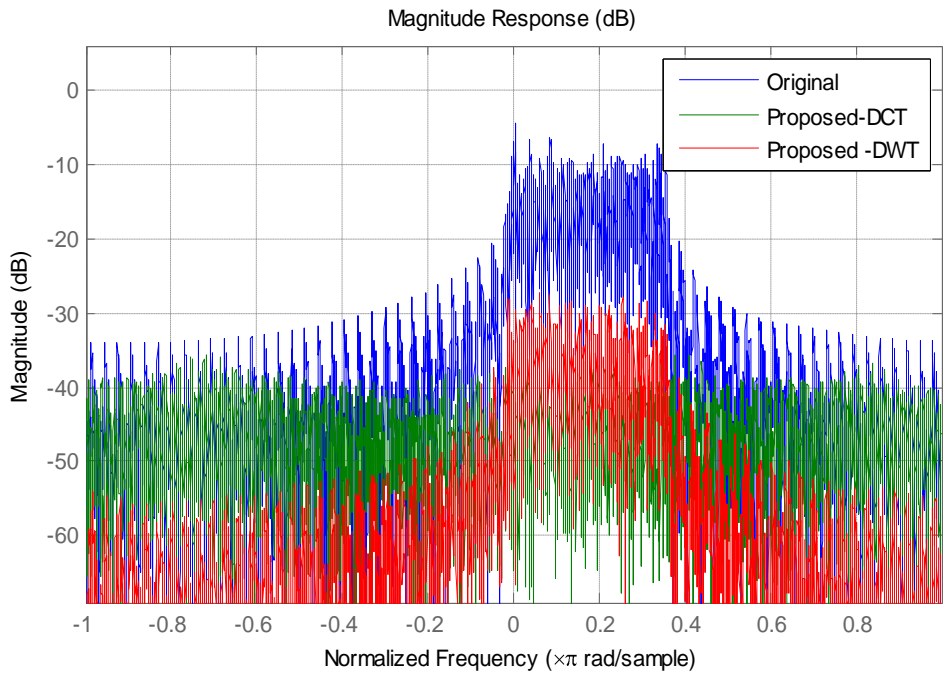

Figure 14. PSD using PN sequence

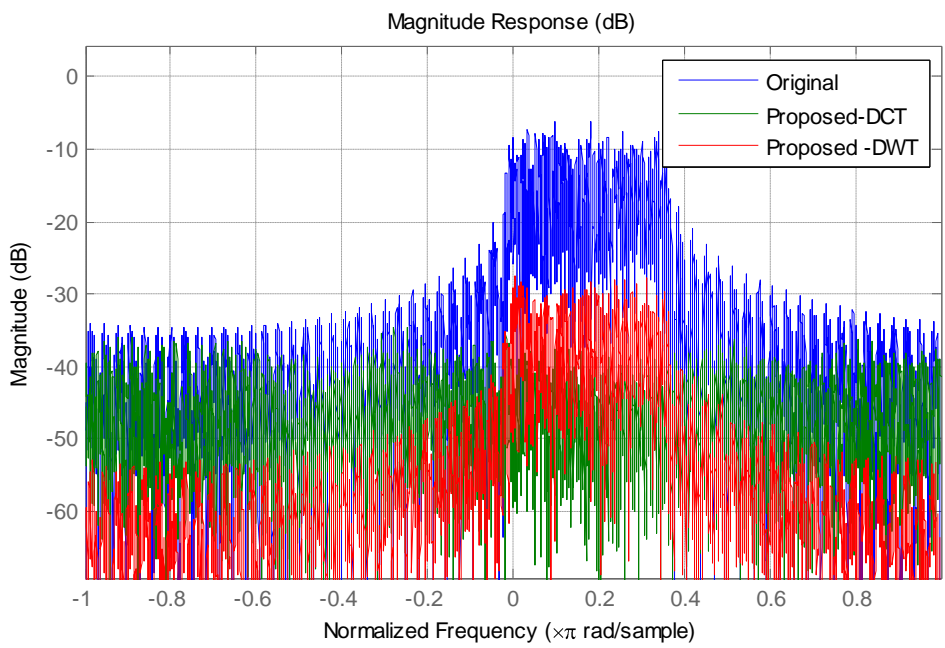

Figure 15. PSD using Gold codes.

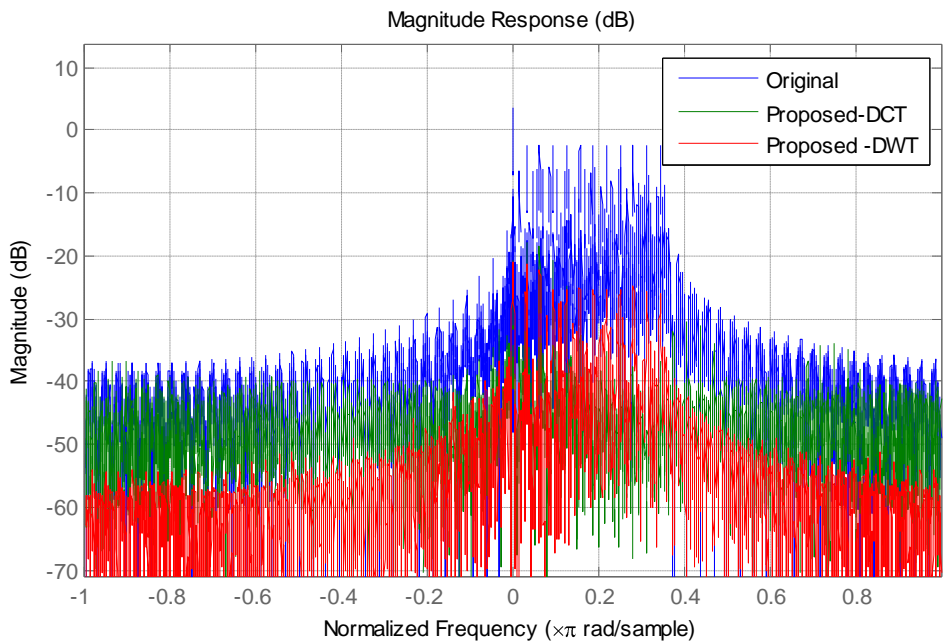

Figure 16. PSD using Walsh Hadamard codes.

The simulation results of Power Spectral Density (PSD) in figures 14, 15, 16 show that the DWT with companding based MC CDMA system has lower side components compared to the original MC CDMA, MC CDMA with companding, and DCT with companding based MC CDMA. DWT with companding based MC CDMA has lower side lobe component, which minimizes interference between signals and has less mean amplitude. It maintains constant main lobe compared to other techniques. Walsh Hadamard code has less mean amplitude in the main lobe when compared with PN sequence and Gold codes. DWT with companding based MC CDMA system is 20 dB to 30dB less in the side lobe component, and 25 dB to 35 dB is less in the main lobe component when compared to other techniques. DCT with companding results in large power variation in the main lobe when compared with the DWT based system.

Table1. Comparison of proposed techniques with other techniques

| **System Type** | **Power increase** | **Implementation complexity** | **BER Degradation** |
| --- | --- | --- | --- |
| Original MC CDMA | Yes | Low | No |
| MC CDMA system with companding | No | High compared to original MC CDMA | No |
| MC CDMA with DCT and companding | No | High | No |
| MC CDMA with DWT and companding | No | High | No |

Table1 shows a comparison between proposed techniques with conventional techniques in terms of power increase, complexity, BER degradation.

## 7. CONCLUSIONS

MC CDMA is used to combat channel distortion, and improves the spectral efficiency, high data rate, robust against multipath fading. In this paper, we implemented a MC CDMA system using combined DCT with companding and DWT with companding to reduce the PAPR. These implemented techniques reduce PAPR more when compared to other conventional MC CDMA techniques. But complexity is increased as DCT algorithm requires $Nlog_2 N$ multiplications and $2\left(\frac{3N}{2}log_2 N - N + 1\right)$ additions which get added in proposed PAPR scheme.

Similarly when the same MC CDMA system is implemented using DWT and companding, the PAPR gets reduced further more when compared to DCT technique. Additionally, the technique is efficient with less distortion and does not require any complex optimization algorithm. The simulation results show that the PAPR reduction is improved by using DWT with companding MC CDMA system compared with DCT and companding MC CDMA system.

The simulation results show that the MC CDMA with DCT and companding scheme obtained about 1 dB PAPR reduction compared with MC CDMA with companding. MC CDMA with DWT and companding method, the PAPR is further reduced by 1dB for µ is 2, 3, and 5 when compared with MC CDMA system with DCT and companding.

The suggested techniques reduce PAPR, and improve the spectrum efficiency. The power spectral density has lower side lobe components and higher bandwidth efficiency compared to DCT and companding.

Proposed system has superior performance in terms of power spectral density and low PAPR. MC CDMA systems' applications are related to personal wireless communication, broadband multi user communications, wireless local area networks and WiMAX broadcasting.

ACKNOWLEDGEMENTS

This work is carried out on the basis of the paper entitled "Combined DCT and Companding for PAPR Reduction in OFDM signals". We wish to offer our sincere gratitude and thanks to Zhongpeng Wang for having motivated us to take up the problem of reducing PAPR in MC CDMA signals.

**Authors**

**Sarala Beeram** has received her B.Tech. & M.Tech. (Digital Systems & Computer Electronics) from Jawaharlal Technological University, Hyderabad in 1993 and 1998 respectively. She is presently working as an Associate Professor in the Department of ECE in M V S R Engineering College, Hyderabad. Her areas of research include CDMA and Multi Carrier CDMA technologies & Wireless Communications. She has presented more than 10 papers in various national & international conferences. She has also published a paper in an international journal.

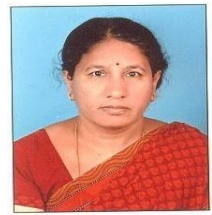

**D.S.Venketeswarlu** received his B.E from Andhra University in 1960, Andhra Pradesh, and M.E from Indian Institute of Science, Bangalore, in 1962, PhD from University of Southampton, UK, in 1967. He has about 13 years industrial experience and about 38 years academic and R&D experience. He has published about 60 papers in peer reviewed journals (National and International) and presented papers in various National and International conferences.

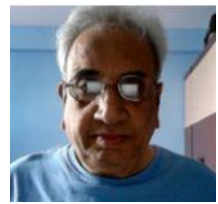